# Ultrafast Imaging Reveals Energy Transport Anisotropy Grows from Bulk to Monolayer Black Phosphorus

*Stephanie M. Hart,[1,#] Siddhartha Sohoni,[2,#] Amal Mathew,[3,4] Aidan L. O'Beirne,[3,5] Bowen Hou,[6] Reef Sweeney[7] Ethan J. Dunsworth,[8] Diana Y. Qiu,[6,9] Tony F. Heinz,[3,4] Naomi S. Ginsberg[1,2,7,10,11,12*]*

1 Department of Chemistry, University of California Berkeley, Berkeley, CA, 94720 USA
2 Molecular Biophysics and Integrated Bioimaging Division, Lawrence Berkeley National Laboratory, Berkeley, CA, 94720 USA
3 Department of Applied Physics, Stanford University, Stanford, CA, 94305, USA
4 SLAC National Accelerator Laboratory, Menlo Park, California 94025, USA
5 Department of Physics, Stanford University, Stanford, CA, 94305, USA
6 Department of Materials Science, Yale University, New Haven, CT, 06520 USA
7 Department of Physics, University of California Berkeley, Berkeley, CA, 94720 USA
8 Engineering Science Program, University of California Berkeley, Berkeley, CA, 94720 USA
9 Energy Science Institute, Yale University, West Haven, CT, 06520 USA
10 Materials Science and Chemical Science Division, Lawrence Berkeley National Lab, Berkeley, CA, 94720 USA
11 Kavli Energy NanoSciences Institute at Berkeley, Berkeley, CA, 94720 USA
12 STROBE, National Science Foundation Science and Technology Center, University of California Berkeley, Berkeley, CA, 94720 USA

# These authors contributed equally to this work.
* Corresponding author, nsginsberg@berkeley.edu

*Abstract:*
Black phosphorus is an emerging low-dimensional semiconductor whose tunable band gap, high carrier mobility, and anisotropic transport make it attractive for directional energy transport in electronics and thermoelectrics. Using ultrafast stroboscopic optical scattering microscopy (stroboSCAT), we image spatiotemporal evolution of energy carriers with picosecond resolution, simultaneously resolving heat and charge transport in bulk black phosphorus and exciton transport in the monolayer. The in-plane armchair:zigzag anisotropy increases from 5.3:1 in the bulk to 6.7:1 in the monolayer, driven by enhanced armchair exciton diffusivity (45.5 $cm^2/s$) relative to bulk armchair charge carriers (16.5 $cm^2/s$). The monolayer anisotropy is smaller than the calculated exciton dispersion alone would suggest, which first-principles calculations attribute to preferential exciton-phonon scattering to large momenta along the zigzag direction. These directly measured, layer-resolved diffusivities identify armchair-aligned monolayers as the optimal geometry for directional transport and show that transport anisotropy cannot be inferred from band dispersion alone.

*Introduction*
Black phosphorus (BP) is a layered elemental semiconductor with a high carrier mobility, tunable bandgap, and strongly anisotropic transport properties. BP has reported carrier mobilities on the order of 500-1000 $cm^2/(V*s)$, exceeding that of transition metal dichalcogenides and approaching that of graphene[1,2]. In contrast to graphene, however, BP has a layer-dependent tunable electronic structure, with an optical gap varying from 0.3 eV in bulk material to ~1.5 eV in the monolayer, owing to the combined effects of quantum confinement and layer-dependent screening[3–6]. Additionally, owing to its puckered crystal structure (Figure 1), BP exhibits layer-dependent in-plane anisotropic transport properties with charge transport enhanced along the armchair direction (AC in Figure 1A, B) and heat transport enhanced in the zigzag direction (ZZ in Figure 1A, B) owing to variations in effective mass and phonon dispersion, respectively[7–11] . This ability to drive directional transport of charge carriers in particular makes BP an intriguing material for field effect transistors. Furthermore, the anisotropies associated with heat transport also point to potential directional heat dissipation properties, as heat management is critical in device operation. Indeed, BP devices have been demonstrated, but largely at the proof of concept level[4,12,13]. In order to develop insight into BP functionality, how transport mechanisms (including both rates and anisotropic characteristics) vary from the bulk to low dimensional limit needs to be clearly depicted. In this work, we investigate these behaviors using ultrafast transient optical scattering microscopy (stroboSCAT).[14–16] Among the ultrafast microscopies that directly image energy carrier transport[15,17–26], stroboSCAT uniquely enables electronic and thermal energy introduced by photoexcitation to be simultaneously spatiotemporally resolved and also distinguished dynamically[14,27–31].

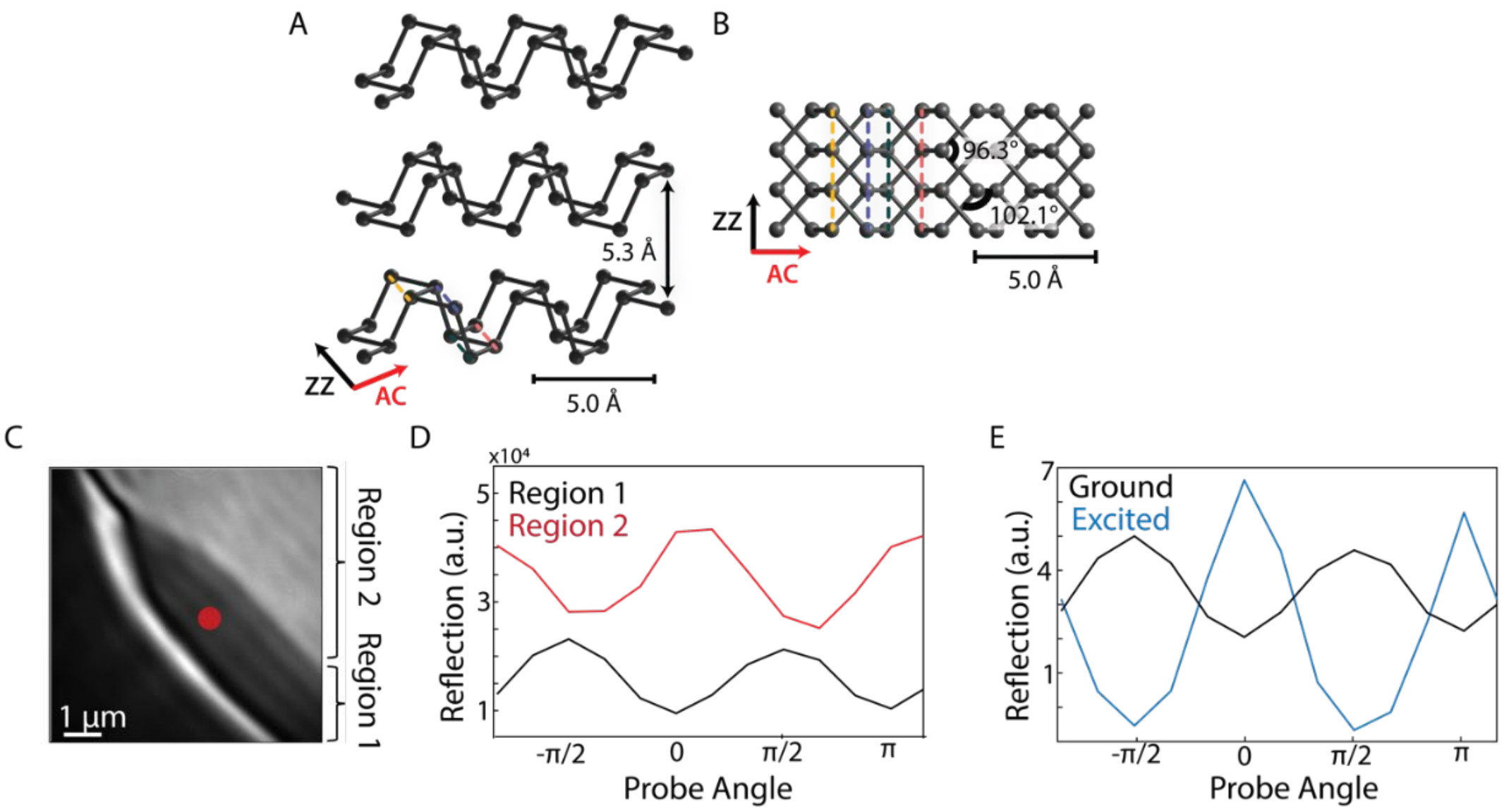

*Figure 1: Bulk black phosphorus structure and birefringence. (A) Side view and (B) top view of black phosphorus crystal lattice.*[32] *(C) Reflectance image of black phosphorus flake. stroboSCAT measurements described in Figure 2 were collected in Region 1 at spot marked with red dot. (D) Ground state reflectance as a function of probe polarization. (E) Region 1 ground state reflectance signal and excited state stroboSCAT signal as a function of probe polarization.*

*Results*

Bulk BP flakes were exfoliated and encapsulated in boron nitride (see **Methods** and Figure S1). Figure 1C shows a ground state reflectance image measured with a 750 nm laser. Variation in the illuminating polarization (Figure 1D) shows how the ground state reflectance changes according to polarization angle ($0^{o}$ was set to the minimum ground state reflectance of Region 1 in Figure 1C). These trends agree with polarization resolved microscopy experiments[33]. In these previous experiments, the minimum in transmission (maximum of reflectance) corresponds to excitation along the armchair direction. Correspondingly with the present results, a probe angle of $0^{o}$ should correspond to excitation along the zigzag direction. Thus, the anticorrelated nature of the angle-dependent reflectance for Region 1 and Region 2 (shown in Figure 1D) suggests that these could be flake regions with orthogonal crystal axes. The excited state reflectance (stroboSCAT) signal ($\Delta R/R$) also displays largely birefringent properties (in line with previous transient reflectance micro-spectroscopy[34]) and is out of phase with the ground state reflectivity intensity as a function of probe polarization.

Time-resolved stroboSCAT images were collected using 600 nm excitation and a 750 nm widefield probe, with an instrument response function of ~0.7 ps. Images were captured with either the pump and probe both polarized along 0$^o$ (Figure 2), or along the 90$^o$ direction (Figure S2) to capture co-polarized dynamics across the two orthogonal crystal axes. Focusing on Figure 2 (0$^o$ data), an initial bright spot is generated by optical excitation that decays on the few picosecond time scale, then giving rise to dark contrast. Owing to the significant separation of time scales of the light and dark contrast decay and diffusion, we assign the light contrast to charge carriers and the long-lived dark contrast to the presence of heat. The source of dark contrast is likely present at early times from above bandgap excitation but is simply overwhelmed by the positive charge carrier signatures.

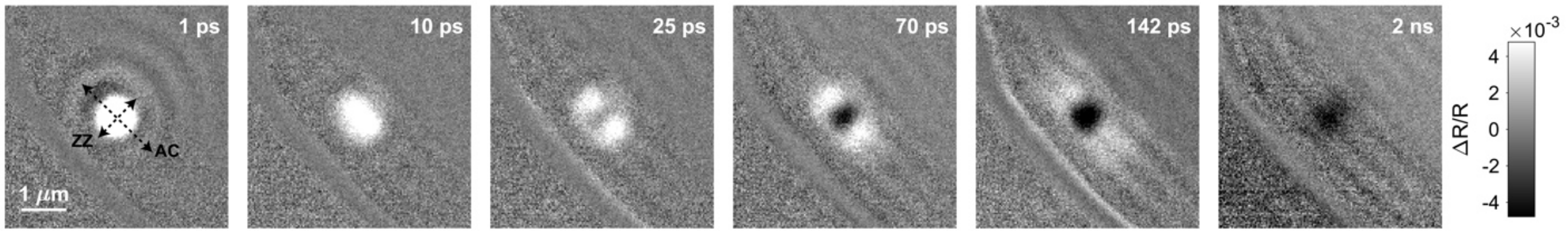

*Figure 2: Bulk black phosphorus stroboSCAT image series collected with co-polarized pump and probe at 0$^o$ (as defined by Figures 1 D,E).*

Integrated population dynamics of the stroboSCAT images are shown in Figure S3A (derived from images shown in Figure 2). The charge carrier population decays on a ~15 ps time scale. Upon subtraction of exponential kinetics, the resultant residual (Figure S3B) shows significant oscillatory features in the transient reflectance signal as a function of time. Fourier transformation of these residual oscillations recovers the spectrum of low frequency acoustic modes (Figure S3C) with a broad band between ~6-30 GHz and a significant peak at ~49 GHz, in line with previous observations[35,36]. The likely mechanism is through a thickness-independent photoelastic modulation (as opposed to thermoelastic modulation). Studies on thickness-dependent recovery of acoustic modes have also suggested that photoelastic coupling to the excitation drives compressive strain responsible for reducing the imaginary index of refraction in the armchair polarization but increasing the real component of the refractive index for the zigzag polarization[37]. Towards recovery of spatially-dependent phonon spectra, we also plot the Fourier transform of residual oscillations as a function of space along the armchair direction in Figure S3D. Even though the positive contrast quickly moves from the original point of photoexcitation (Figure 2, and further discussed in reference to Figure 3), the power spectra only show signatures around the initial ~300 nm confocal excitation spot size. This further points to a photoelastic response owing to impulsive excitation, similar to impulsive stimulated Raman scattering. Furthermore, the spectra show minimal evolution as a function of space even within the ~300 nm associated with the photoexcitation spot size.

In addition to the evolution from light to dark contrast shown in Figure 2, clear spatial evolution is shown as a function of time. The light contrast associated with charge carriers expands more rapidly along the antidiagonal direction, beginning from a round spot at 1 ps, and resulting in an oval by 10 and 25 ps. This evolution suggests that this antidiagonal direction is the armchair direction (noted AC in the 1 ps image). Additionally, the dark contrast signature associated with heat expands more rapidly along the diagonal direction, pointing to its assignment to the zigzag (ZZ) axis, also marked in the 1 ps image. These trends are further quantified by plotting the mean squared expansion as a function of time along each of these axes, as shown in Figure 3. After isolating the trend from the photoelastic modulation (see SI and Figures

S4-S9), the diffusion coefficients recovered for the charge carrier contrast are 16.5 $cm^2/s$ and 3.0 $cm^2/s$ for the armchair and zigzag axes, respectively, yielding a 5.5:1 anisotropy ratio for these two directions. A similar analysis targeting the heat contrast recovers diffusion coefficients on the order of ~1 $cm^2/s$ along both axes (Figures S10-12), but the uncertainty precludes obtaining a reliable anisotropy ratio.

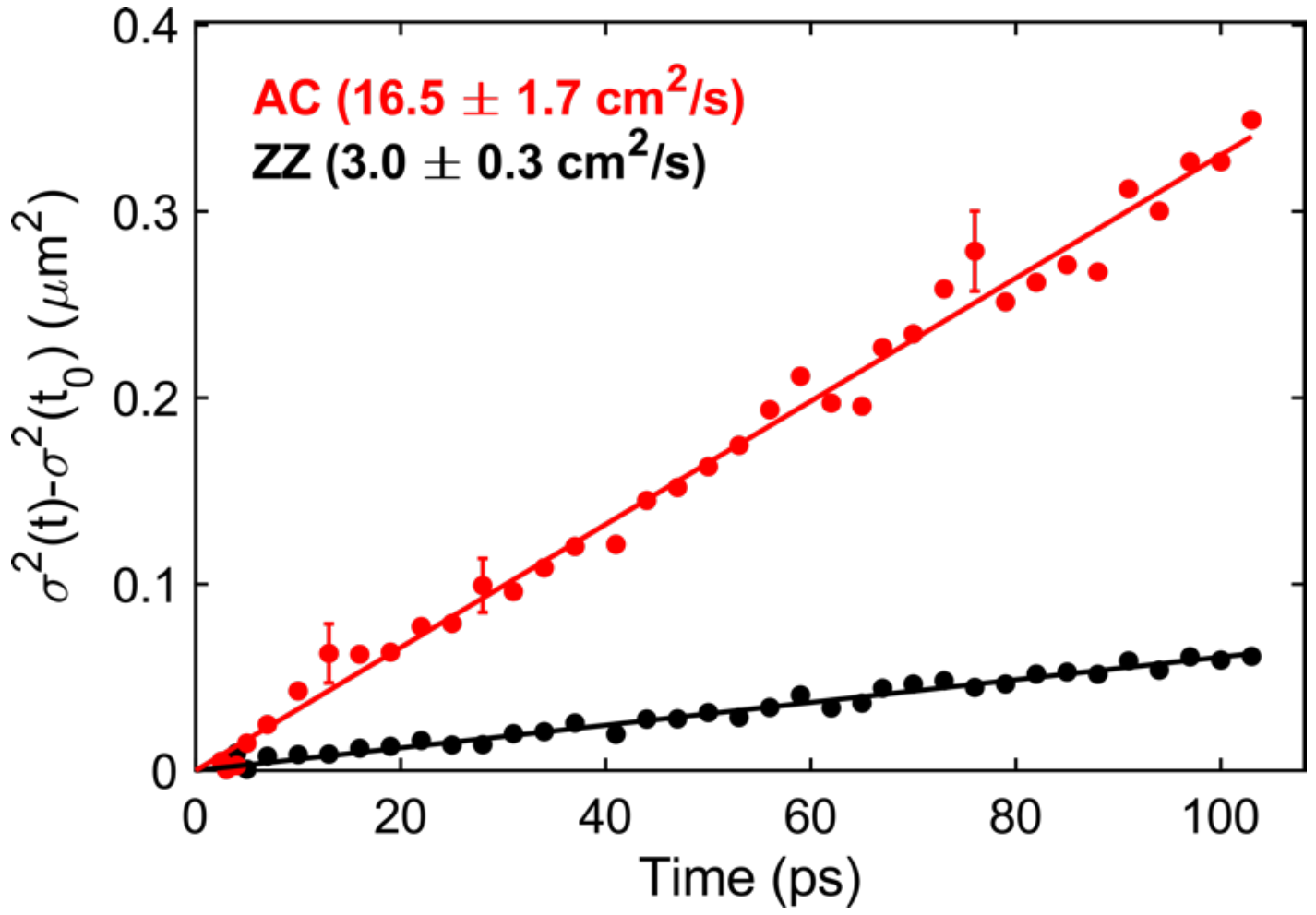

*Figure 3: stroboSCAT-resolved bulk electronic transport characteristics. (A) Mean squared expansion of charge carrier contrast as a function of time along armchair (AC) and zigzag (ZZ) directions (co-polarized pump and probe at 0°). Diffusion coefficient (in legend) derived from assuming a normally diffusive (linear) system.*

While the bulk transport characteristics have been examined by transient reflectance and ultrafast electron microscopy measurements, the spatially resolved comparison to the few-layer and monolayer limit has been lacking to-date. To this end, monolayer regions of BP flakes were identified by their optical emission signatures and were correlated to individual flakes using a combination of photoluminescence and white light microscopy (Figure S13). Time-resolved stroboSCAT images recovered from photoexcitation of the monolayer are shown in Figure 4A. The co-polarized pump and probe were set to maximize the time-zero excited state signal. Here the excitons generated immediately upon photoexcitation give rise to dark contrast, while heat signatures are present as light contrast. The minimal heat signatures relative to the bulk likely arise from the excitation at 600 nm being much closer to the ~1.5 eV bandgap (as compared to the ~0.3 eV bulk bandgap). The variation in overall contrast sign as compared to the bulk material could result from (1) probing a different region relative to the bandgap, (2) thickness variation, or (3) BP crystalline axis orientation. The transport characteristics across the armchair and zigzag directions were extracted for the photogenerated exciton in the same fashion as for the charge carriers in the bulk material (heat signatures were too minimal to quantify). We observe enhanced excited-state mobility in the armchair direction, with a diffusion coefficient of 45.5 $cm^2/s$ (Figure S14), while the diffusion coefficient in the zigzag direction is 6.9 $cm^2/s$ (Figure S15), comparable to the corresponding bulk value. Together, these observations yield an anisotropic transport ratio of 6.7:1, moderately larger than that obtained in bulk.

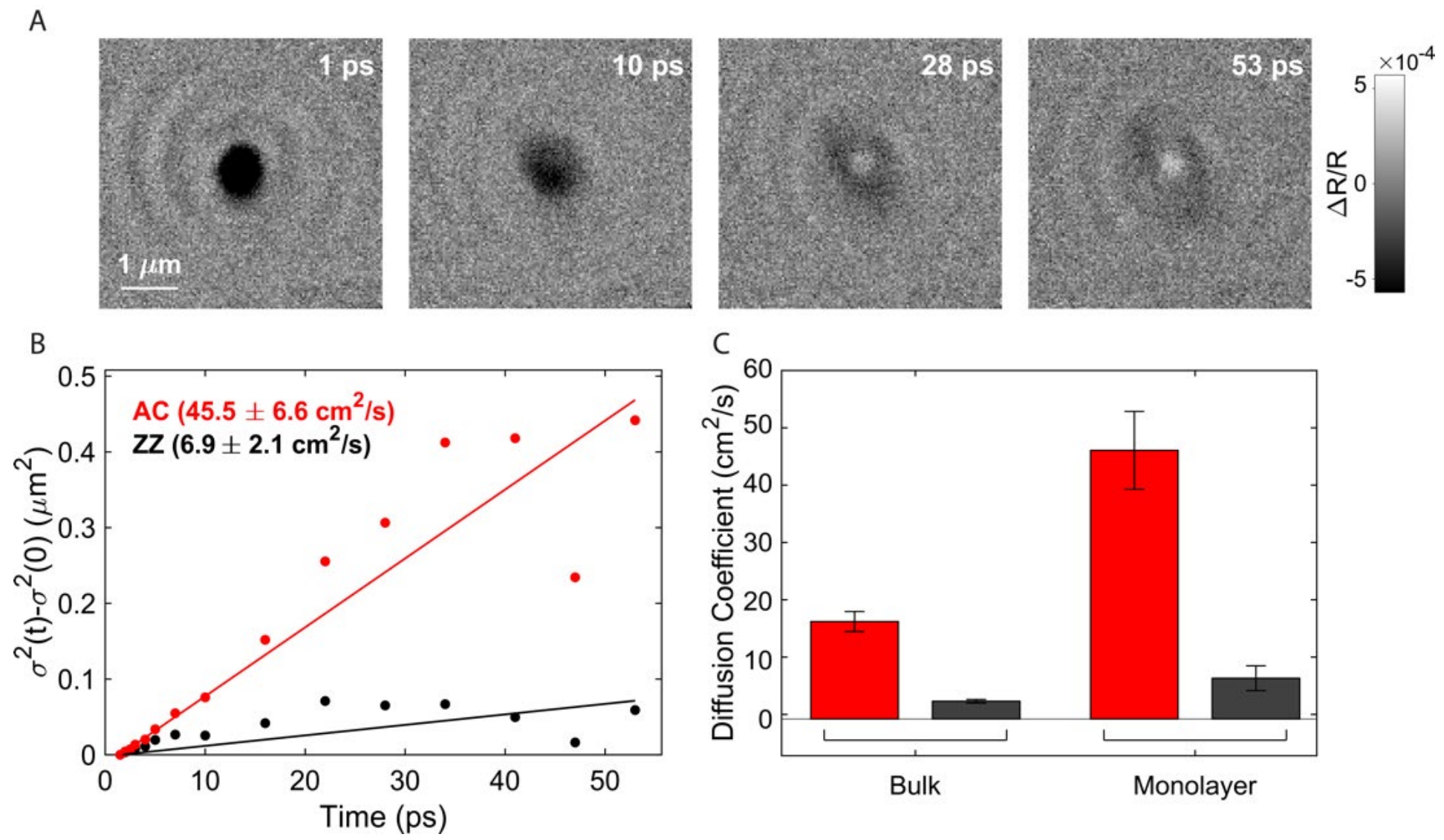

*Figure 4: Monolayer black phosphorus exciton transport. (A) stroboSCAT images collected for co-polarized pump and probe to maximize excited state contrast (equivalent to 0° in Figure 1 D,E). (B) Mean squared expansion as a function of time for exciton contrast along the armchair (AC) and zigzag directions used to derive diffusion coefficients (shown in legend). (C) Comparison of charge-carrier and exciton diffusion for bulk and monolayer black phosphorus. The monolayer displays an enhancement in excited state transport along with a moderate increase in transport anisotropy between the armchair and zigzag axes, relative to the bulk.*

Previous GW + Bethe-Salpeter equation (GW-BSE) calculations of the exciton bandstructure of black phosphorus predicted a strongly anisotropic exciton dispersion, with a much lighter effective mass along the armchair direction than along the zigzag direction.[38] To understand the experimentally measured diffusion, we perform first principles calculations of the exciton dispersion[39], exciton-phonon scattering rates[40–43], and exciton transport within a Boltzmann transport equation approach[43–45]. We start by calculating excitons in monolayer black phosphorus within an ab initio GW-BSE formalism[46–48] and then simulate the time evolution of an excited exciton wavepacket with a spread of 8 Å centered at the lowest energy bright exciton within a Boltzmann transport equation (BTE) approach with exciton-phonon interactions calculated within many-body perturbation theory (MBPT) [40,41] included in the collision matrix. Further details can be found in the SI. Both the quasiparticle band structure and the exciton band structure (Figure 5A,B) show a strong anisotropy. In the exciton bandstructure, the long-range exchange interaction gives rise to a linear dispersion along the dipole-allowed armchair direction, in the limit of zero exciton center of mass momentum ($Q \to 0$), and a parabolic dispersion along the dipole forbidden zigzag direction, consistent with previous results[38]. The exciton effective mass is 4.16$m_0$ along the zigzag direction and approximately 0.53$m_0$ along the armchair direction if we fit the effective mass outside the linear dispersion regime (here, $m_0$ is the bare electron mass). Interestingly, even though the exciton disperses more along the armchair ($\Gamma - X$) direction, when exciton-phonon interactions are taken into account, the exciton is more likely to scatter in the zigzag direction ($\Gamma - Y$) because the slope of the exciton dispersion in the armchair direction is steeper than the slope of the acoustic phonon dispersion. The inset of Figure 5B illustrates the exciton-phonon scattering rate for the lowest-energy bright exciton (indicated by the arrow) into all other exciton states in the Brillouin zone. Due to the anisotropy of the bandstructure, the exciton is able to scatter to much larger momenta along the zigzag ($\Gamma - Y$) direction.

We model the transport anisotropy by exciting an exciton wavepacket and modeling its evolution in time using the BTE Fourier transformed to real space (Figure 5C). The simulated process is not a full description of diffusion, since it only includes exciton-phonon scattering and neglects exciton population effects, but it provides qualitative insight into how the anisotropic exciton-phonon coupling and exciton dispersion manifest in the exciton transport. In this approach, we see a clear transport anisotropy, and the ratio of the mean spread of the exciton wavepacket in time in the $X$ and $Y$ directions is $\frac{\langle r_x(t)\rangle}{\langle r_y(t)\rangle} \approx 2.46$. Within the relaxation time approximation, the diffusion coefficient in 2D is $D = \frac{1}{2}\langle v^2\rangle\tau$.[49] Here, $\tau$ is the relaxation time and $v$ is the group velocity associated with the band curvature. We approximate the diffusion coefficient by extracting the exciton group velocities at different center-of-mass momenta from the calculated exciton bandstructure and estimate a relaxation time of 25 fs at room temperature by considering the exciton-phonon lifetime for each exciton state, since different excitons in the bandstructure have significantly different lifetimes. This results in an estimate of the diffusion coefficient for each finite momentum exciton state as shown in Figure 5D. Comparing to the experimentally obtained exciton diffusivity, this estimate achieves reasonable order-of-magnitude agreement of tens of $cm^2/s$ along the armchair axis (peak computed diffusivity of 27 $cm^2/s$, compared to 45.5 $cm^2/s$ obtained experimentally) and also reasonable order of magnitude agreement along the zigzag axis, with computed values ranging between 0.01- 2.2 $cm^2/s$ relative to a measured value of 6.9 $cm^2/s$, suggesting that the exciton-phonon scattering is sufficient to account for the transport anisotropy seen in experiment.

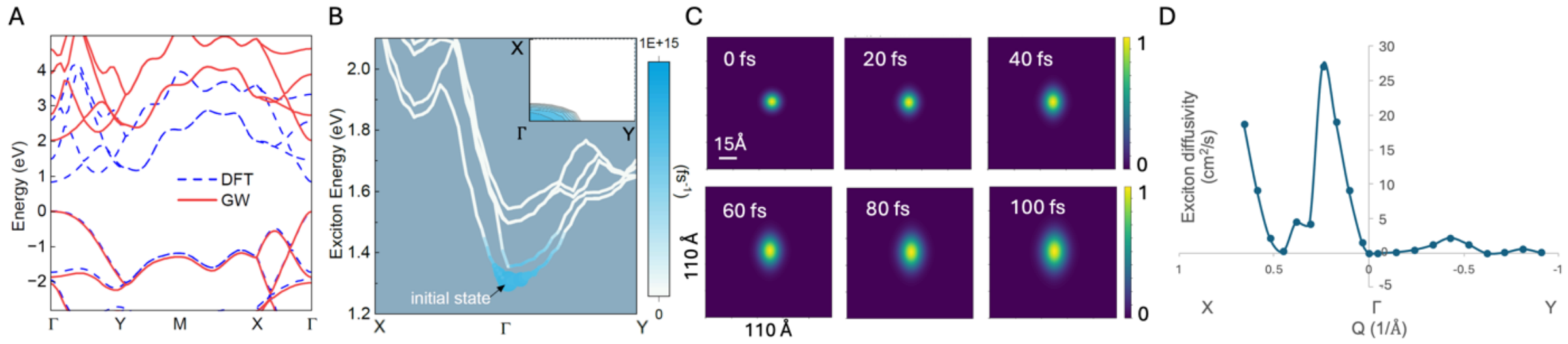

*Figure 5. (A) DFT (blue dashed curve) and GW bandstructures (solid red curve). (B) Exciton band structure colored by exciton-phonon scattering rate and with an arrow at the initial state for the BTE. The inset shows the contour line of the exciton-phonon scattering rate of the initial state in the Q space. (C) Real-space BTE image from t=0 to 100 fs. (D) Exciton diffusivity of the first exciton band.*

*Discussion*

The order-of-magnitude agreement between our experimentally measured and theoretically predicted transport parameters for monolayer BP, together with the first-principles account of the transport anisotropy, points to a consistent picture in which exciton-phonon scattering reduces the transport anisotropy relative to the anisotropy of the band structure. They furthermore suggest that the strategy that we employed to obtain bulk values from stroboSCAT experiments should be relatively accurate as well. Diffusion coefficients for electronic excited states for the armchair and zigzag axes have been previously measured by both ultrafast transmission electron microscopy (TEM) and transient reflectance microscopy (and correspond primarily to hole transport) [50,51]. Armchair diffusivity values in the range of ~1300 $cm^2/s$ have been reported along with zigzag diffusivities ranging from 80-800 $cm^2/s$ [8,9]. As such, both the overall diffusion coefficients obtained from stroboSCAT and the anisotropic ratio between the armchair and zigzag axes are lower than previously reported values. To date, it is unclear if this wide range of results is a function of time resolution, excitation density, or sample preparation procedure (e.g., defect density). Presumably, differences in experimental methods can come into play, including how data are modeled, fit, and interpreted, especially in the way that thermal signatures are/not distinguished from electronic ones[28] which we carefully describe in the SI, and the extent to which excitation densities promote higher order carrier-carrier interactions. Previous work has established a heat diffusivity on the order of 0.1-0.4 $cm^2/s$ with a ~2:1 anisotropy ratio favoring the zigzag axes that decreases for thinner films[10,11]. To within our uncertainty, our results are consistent with this previous work.

The electronic transport in both the bulk material and the monolayer (charge carriers and excitons, respectively) provide a useful comparison in their own right. Here, the transport anisotropy ratio between the armchair and zigzag axes changes with layer thickness, going from 5.3:1 in bulk to 6.7:1 in the monolayer. The increase in this anisotropy in the monolayer is primarily due to the strong increase in armchair direction exciton diffusivity of 45.5 $cm^2/s$ relative to the ambipolar transport of electrons and holes in the bulk of 16.5 $cm^2/s$. Naively, one might expect that the thickness dependence decreases at the monolayer level by favoring free carriers to bound excitons, or that the scattering time or exciton or defect densities could play a role. Although Boltzmann transport calculations for the bulk material are currently prohibitive, intuition from the layer dependence of the excitonic/electronic band structure calculations[38] support the observed enhancement: while the dispersion does become shallower as the number of layers is decreased to a single one, the range of momenta near the Γ point where this is true becomes smaller. This raises the interesting question of how to relate the dispersion in mobility to experimental observations, which presumably include carriers with multiple momenta. As we accumulate additional combined experimental/computational transport results on additional systems in the future, it should be possible to develop more explicit such relationships, although the observed agreement between theory and experiment here suggests that the current strategy that we have employed is generally acceptable.

Beyond the technical and scientific details, this work affirms that monolayer black phosphorus offers the possibility of high-mobility and directional transport, which could be highly useful for next-generation field effect transistors. Furthermore, the anisotropy in electronic and thermal transport could be leveraged not only for thermal management in devices but also for thermoelectric applications in which waste heat can be converted back to useful work. Diagnostic tools like stroboSCAT, especially in combination with advanced electronic structure and dynamics calculations, to incisively characterize the potential for such effects will be important not only for BP but also for a range of other materials in which heat and charge excitations manifest prominently and need to be spatiotemporally distinguished and resolved. For example, other thermoelectric van der Waals materials[52] or those currently under development like doped $Ag_2Se$[53] could be more clearly investigated with a similar approach.

*Acknowledgements*
Sample fabrication and characterization was supported at SLAC by the AMOS program, Chemical Sciences, Geosciences, and Biosciences Division, Basic Energy Sciences (BES), Office of Science (SC), US Department of Energy (DOE) and at Stanford by the National Science Foundation (grant EFMA 2425218). Measurement and theory were supported by the Center for Computational Study of Excited State Phenomena in Energy Materials (grant no. C2SEPEM) under the US Department of Energy, Office of Science, Basic Energy Sciences, Materials Sciences and Engineering Division (contract no. DE-AC02-05CH11231), as part of the Computational Materials Sciences Program. Instrumentation and methodology were supported by the STROBE Center for Realtime Imaging, a National Science Foundation Science and Technology Center (grant DMR 1548924). Data analysis was supported by the Director, Office of Science, Chemical Sciences, Geosciences, and Biosciences Division, of the U.S. Department of Energy under Contract no. DEAC02-05CH11231. S.M.H. was supported by an Arnold O. Beckman Postdoctoral Fellowship in Chemical Sciences. N.S.G. and D.Y.Q. were also supported by David and Lucile Packard Foundation Fellowships for Science and Engineering. The calculations used resources of the National Energy Research Scientific Computing (NERSC), a DOE Office of Science User Facility operated under contract no. DE-AC02-05CH11231, under award BES-ERCAP-0031507 and BES-ERCAP-0027380; and the Texas Advanced Computing Center (TACC) at The University of Texas at Austin.

Supporting Information Available:
Materials and methods including sample preparation, stroboSCAT setup, ground state reflectance measurements and polarization control, data analysis, and details of Boltzmann transport calculations.

Notes: The Authors declare no competing financial interest.

**Supporting Information for**

# Ultrafast Imaging Reveals Energy Transport Anisotropy Grows from Bulk to Monolayer Black Phosphorus

*Stephanie M. Hart,[1,#] Siddhartha Sohoni,[2,#] Amal Mathew,[3,4] Aidan L. O'Beirne,[3,5] Bowen Hou,[6] Reef Sweeney[7] Ethan J. Dunsworth,[8] Diana Y. Qiu,[6,9] Tony F. Heinz,[3,4] Naomi S. Ginsberg[1,2,7,10,11,12*]*

1 Department of Chemistry, University of California Berkeley, Berkeley, CA, 94720 USA
2 Molecular Biophysics and Integrated Bioimaging Division, Lawrence Berkeley National Laboratory, Berkeley, CA, 94720 USA
3 Department of Applied Physics, Stanford University, Stanford, CA, 94305, USA
4 SLAC National Accelerator Laboratory, Menlo Park, California 94025, USA
5 Department of Physics, Stanford University, Stanford, CA, 94305, USA
6 Department of Materials Science, Yale University, New Haven, CT, 06520 USA
7 Department of Physics, University of California Berkeley, Berkeley, CA, 94720 USA
8 Engineering Science Program, University of California Berkeley, Berkeley, CA, 94720 USA
9 Energy Science Institute, Yale University, West Haven, CT, 06520 USA
10 Materials Science and Chemical Science Division, Lawrence Berkeley National Lab, Berkeley, CA, 94720 USA
11 Kavli Energy NanoSciences Institute at Berkeley, Berkeley, CA, 94720 USA
12 STROBE, National Science Foundation Science and Technology Center, University of California Berkeley, Berkeley, CA, 94720 USA

# These authors contributed equally to this work.
* Corresponding author, nsginsberg@berkeley.edu

# Methods

**Sample preparation**

Black phosphorus (BP) is an extremely air sensitive material with bare samples oxidizing in 20 minutes under ambient atmospheric conditions. To prevent oxidation, both monolayer and bulk BP flakes are exfoliated and encapsulated with hexagonal boron nitride (hBN), in a nitrogen-filled glovebox (MBraun Unilab pro sp) with $O_2$ concentration < 2 ppm and $H_2O$ concentration < 1 ppm.

We first transfer a thin flake of hBN (~ 50 nm) onto a 0.15 mm thick borosilicate glass substrate (VWR VistaVision No. 1.5 Square Cover Glasses). The substrate is then annealed in the glovebox at 500°C to remove adsorbed oxygen and water on the hBN surface.

We then exfoliate BP via standard scotch-tape exfoliation techniques onto a PDMS substrate (3 cm × 3 cm) placed on a glass slide. Monolayer BP flakes on the substrate are then located using a home-built automated scanning system as follows: The substrate is illuminated by a 470 nm light source and then imaged through a 760 ± 10 nm bandpass filter to identify fluorescence of monolayer BP flakes. An automated system scans through the area of the substrate and locates possible monolayer flakes by this method. For both monolayer and bulk flakes, the PDMS substrate is then cut around the location of a selected flake to prepare a stamp of area 5 mm × 5 mm. The flake is then transferred onto the prepared substrate of hBN on cover glass and subsequently encapsulated with another layer of hBN to prevent degradation.

A rubber O-ring (inner diameter 1 cm) is then glued onto the substrate, encircling the sample, and capped by another square cover glass. This is done to further isolate the sample from ambient air conditions during transport and measurements. The sample is measured through the glass on which the encapsulated BP stack rests (**Figure S1**).

Samples made following this procedure have been characterized by absorption and photoluminescence measurements. They have shown narrow excitonic features (~ 15 meV) at low temperatures (~ 4 K) soon after fabrication[1].

**stroboSCAT microscopy**

The stroboSCAT setup has been described in detail previously[2–4]. Briefly, a 630 nm pump pulse was generated by a Light Conversion non-collinear optical parametric amplifier (model PN13F1) using the third harmonic of the fundamental 1080 nm seed pulse from a Light Conversion Pharos laser (model 10-200-PP). A 730 nm probe pulse was generated by a Light Conversion non-collinear optical parametric amplifier (model PN15F2) using the second harmonic of the fundamental seed pulse. The laser repetition rate was 200 kHz; the pump was modulated at 660 Hz, and the pump-probe delay times were controlled by a mechanical stage.

The pump and probe beams were spatially filtered through 25 μm and 50 μm pinholes, respectively. The pump beam was telescoped to a diameter of 12 mm, while the probe beam was telescoped to 1 mm and focused into the back focal plane of the objective of a home-built microscope using an $f$ = 300 mm lens. The two beams were combined using a dichroic mirror (DMLP650, Thorlabs), and a 50/50 beamsplitter reflected both beams into a high numerical aperture (1.4 NA) oil-immersion objective (Leica, 1.40 NA) and onto the sample, resulting in overlapped confocal pump and wide-field probe illumination. Probe light reflected from the sample-substrate interface, as well as the light scattered from the sample, was collected through the same objective. The probe light was isolated with a long-pass filter (FEL0700, Thorlabs) and focused onto a charged metal oxide semiconductor (CMOS) detector triggered at 660 Hz (PixeLINK PL-D752, equipped with a Sony IMX174 global shutter sensor), using an $f$ = 500 mm lens placed one tube length (200 mm) away from the back focal plane of the objective. The total magnification is calculated as 157.5×, giving a spatial resolution of 37.2 nm/pixel. Data were cropped to maintain a field of view of 4 µm by 4 µm. The transient reflectance signal was generated by taking the difference in integrated signal between pump-on and pump-off frames, normalized to the integrated pump-off intensities, yielding $\Delta R/R$. Setup automation and data acquisition were implemented in LabVIEW 2014 (64-bit). Data analysis and plotting were performed using a combination of ImageJ (Fiji) and Matlab. A pump fluence of 36 mJ/cm$^2$ was used.

### Ground state reflectance measurements and polarization control

Ground state reflectance measurements were performed in the stroboSCAT setup described above by blocking the pump beam. A waveplate followed by a polarizer was used to vary the polarization of the probe beam with respect to the sample.

### Ab Initio GW-BSE calculations

We start from a density functional theory calculation using the Quantum Espresso software package[5] The calculation is performed on a black phosphorus monolayer with a lattice constant of 3.295486 Å along $x$ and 4.543635 Å along $y$. The DFT calculations used norm-conserving pseudopotentials with an PBE exchange correlation functional[6]. A wavefunction cutoff of 75 Ry was used for the planewave basis, and a $k$-grid of 12 × 12 × 1 was used to converge the self-consistent charge density.

A one-shot GW calculation using the BerkeleyGW software package[7] was then used to obtain the quasiparticle bandstructure. A kinetic energy cutoff of 25 Ry was used to truncate the planewave basis for the screened Coulomb interaction. The dielectric matrix was calculated on a uniform grid of 12 × 12 × 1 $q$-points. Additionally, 10 subsampled points in the Voronoi cell around $q = 0$ were used to speed up converge with respect to $k$-point sampling, following the nonuniform neck subsampling (nns) scheme. Stochastic subsampling was used to accelerate the sum over empty states[8]. The frequency-dependence was included via the Hybertsen-Louie Generalized Plasmon Pole model (HL-GPP)[9]. The Coulomb interaction is truncated in the aperiodic direction to prevent spurious interactions between periodic images[10] Finite momentum BSE calculations[11] were then used to obtain the exciton bandstructure.

### Boltzmann transport calculations

The coupled real- and momentum-space dynamics of photoexcited excitons is described by the semiclassical Boltzmann transport equation[12] for the exciton occupation $n_S(\boldsymbol{Q},\boldsymbol{R})$ of exciton band $S$ with center-of-mass momentum $\boldsymbol{Q}$ at position $\boldsymbol{R}$:

$$\frac{\partial n_S(\boldsymbol{Q},\boldsymbol{R})}{\partial t} = -\frac{\partial n_S(\boldsymbol{Q},\boldsymbol{R})}{\partial x}\frac{\partial \Omega_S}{\partial Q_x} - \frac{\partial n_S(\boldsymbol{Q},\boldsymbol{R})}{\partial y}\frac{\partial \Omega_S}{\partial Q_y} + K_{\text{scat}}[n_S(\boldsymbol{Q},\boldsymbol{R})]. \qquad \textit{Equation S1}$$

The first two terms describe the ballistic propagation of the exciton wave packet, governed by the band velocities $\partial\Omega_S/\,\partial Q_{x,y}$ obtained from the exciton band dispersion $\Omega_S(\boldsymbol{Q})$, and effectively couple the spatial and momentum coordinates. The last term, $K_{\text{scat}}$, collects the phonon-induced scattering processes, i.e., scattering into and out of each momentum state $\boldsymbol{Q}$, with rates computed from the first-principles exciton–phonon coupling matrix elements via Fermi's golden rule, including phonon absorption and emission weighted by the Bose–Einstein occupations[12].

Equation S1 is propagated numerically on discrete $\boldsymbol{Q}$ and $\boldsymbol{R}$ grids, starting from a Gaussian wave packet in both momentum and real space that mimics the initial laser excitation. The solution yields the time-dependent exciton population and spatial distribution, from which the mean squared displacement and diffusion coefficients are extracted.

# Supporting text and figures

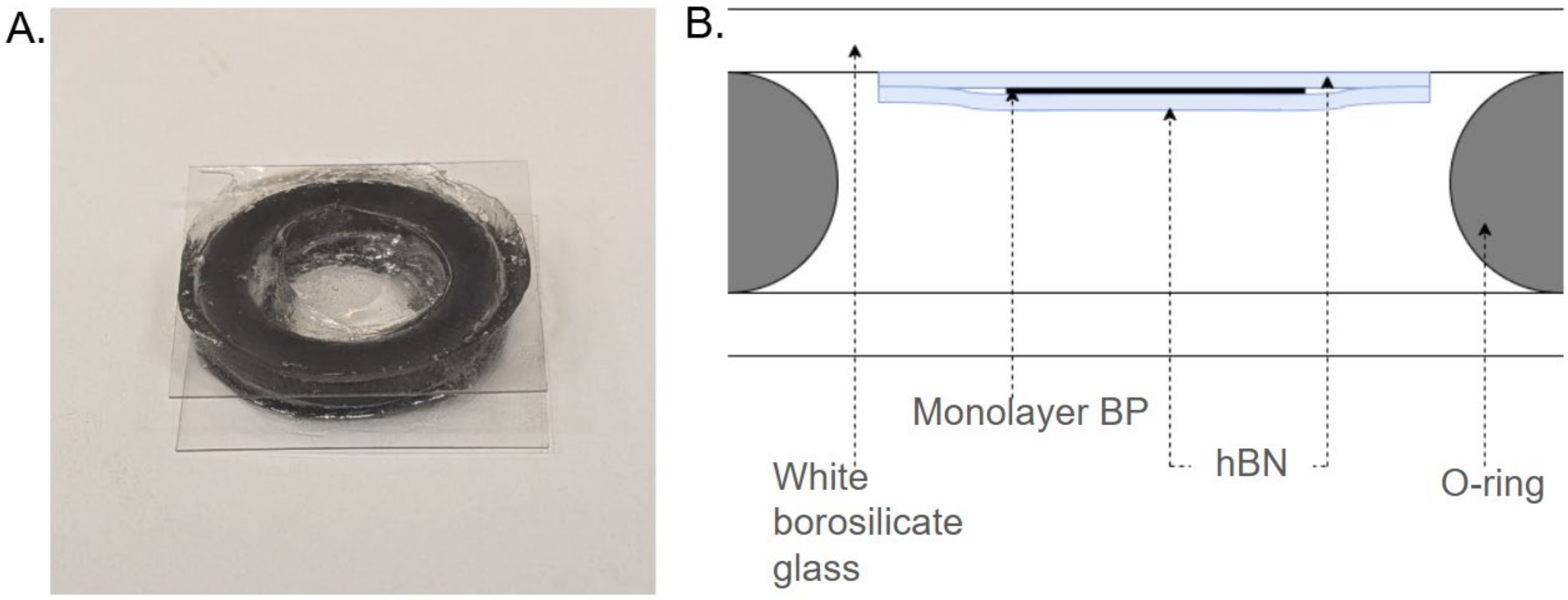

**Figure S1**: Sample preparation. (A) Image of a final sample stack after preparation (B) Schematic of vertical cross section of the sample stack (not to scale).

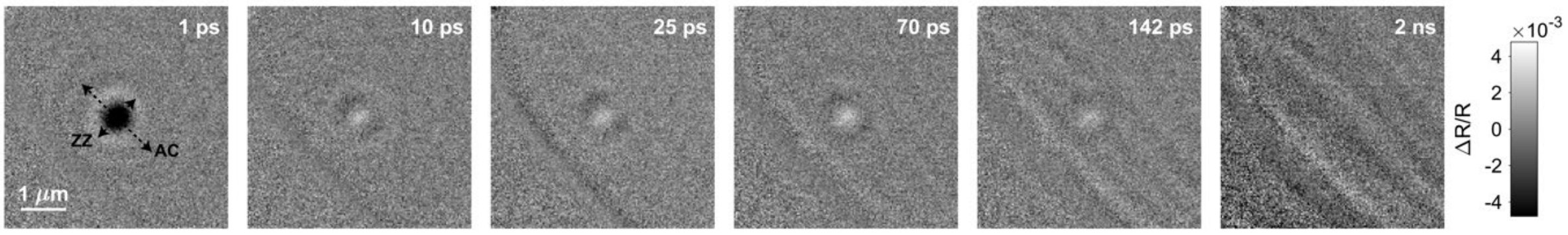

**Figure S2**: Bulk black phosphorus stroboSCAT image series collected with co-polarized pump and probe at 90° (as defined by Figures 1D,E).

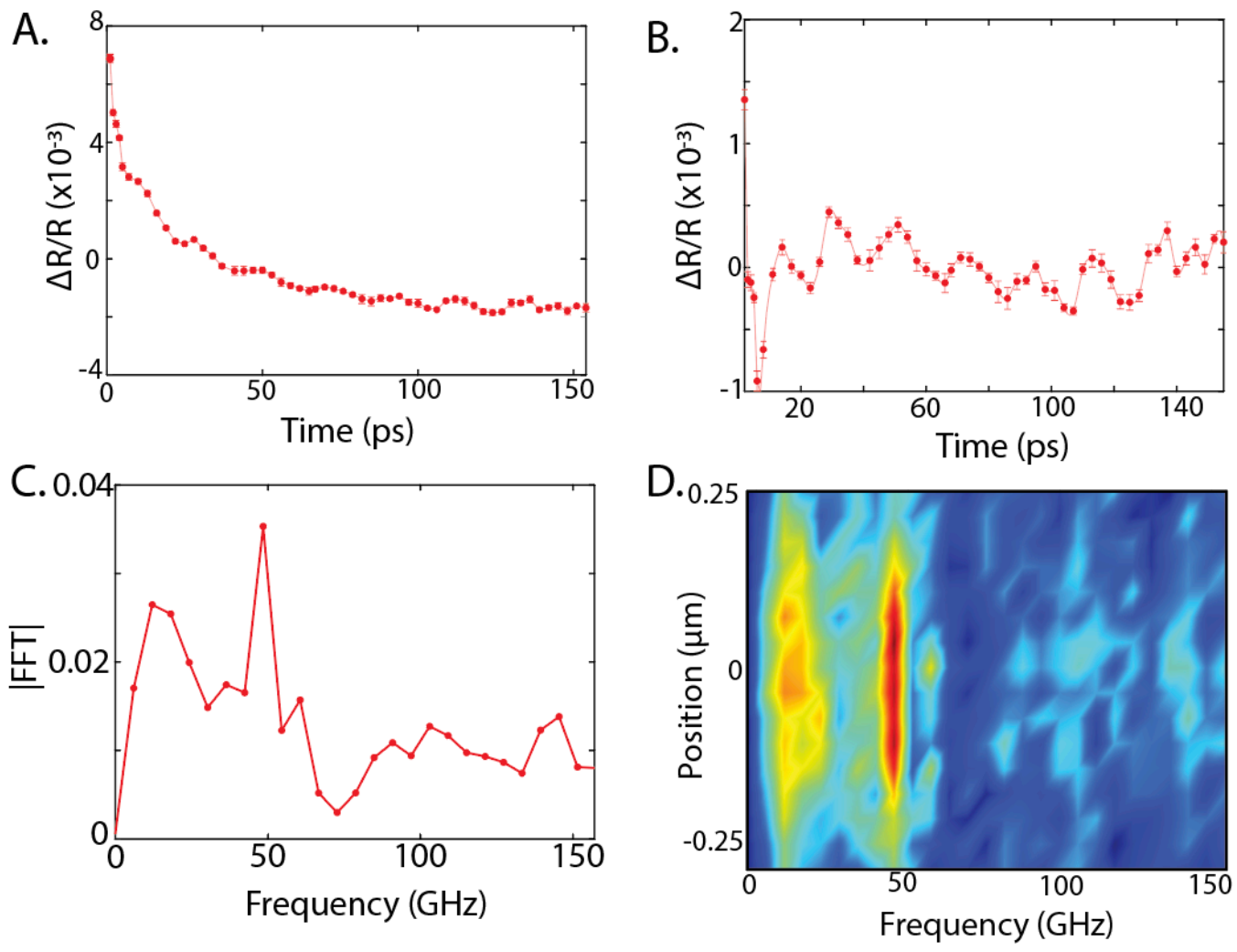

**Figure S3**: Integrated transient reflectance signal reveals acoustic oscillations. (A) Integrated intensity of stroboSCAT signal at 0° pump/probe (data from Figure 2). (B) residual oscillations remaining from panel (A) kinetics after subtraction of exponential decay. (C) Power spectrum derived from Fourier transform of trace in panel (B) showing signatures of low frequency acoustic modes at 49 and ~6-30 GHz. (D) Spatially resolved phonon spectrum within excited spot size.

**Data analysis: electronic diffusivity calculation**

The stroboSCAT data were assembled into a three-dimensional cube (*spatial x * spatial y * delay time)*. In each time delay frame in this cube, the spatially anisotropic signal is described by the sum of two two-dimensional Gaussians: a broader component assigned to the electronic signal (positive contrast in bulk, negative in monolayer) and a narrower opposite-signed component (heat, see Figure 2 in main text). To extract the electronic diffusivity from the stroboSCAT data cube, the contribution of the heat is first masked out by excluding a central elliptical region of data points in the image at each delay time before fitting to a single two-dimensional Gaussian to extract the width of the electronic species distribution.

*Validation of the mask-based electronic width recovery*: Central masking of data points is required to obtain the width of the electronic distribution because the superposition with the narrower, opposite-signed (heat) feature yields a signal profile with a flattened center (**Figure S4**), biasing single-Gaussian fits when the center is retained as shown in **Figure S5**. This behavior was validated using a simple model consisting of the sum of two one-dimensional Gaussians with different widths and opposite signs. Single-Gaussian fits to the combined signal recover the outer (electronic) width only in the large-mask regime, i.e., when sufficient central points are masked (**Figures S5 and S6)**.

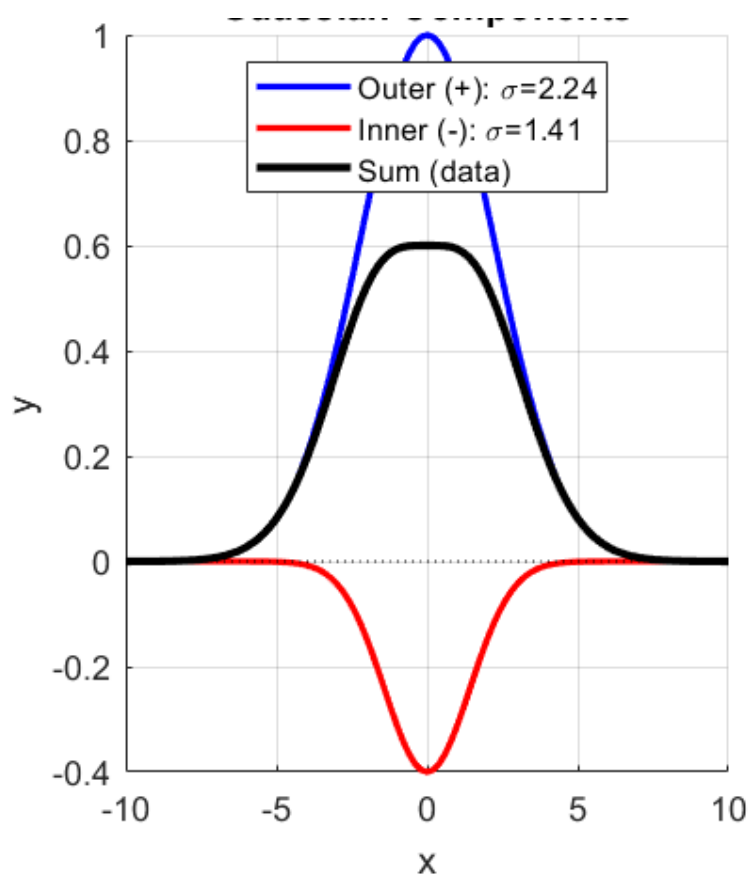

**Figure S4:** Example model of a sum of two one-dimensional Gaussians. A broad positive Gaussian with width 2.24 (blue) and a smaller negative Gaussian with width 1.41 are added together to create the curve shown in black. The resulting black curve is used to demonstrate fitting procedures used to obtain diffusivity from the experimental data in **Figures S7 and S8** below.

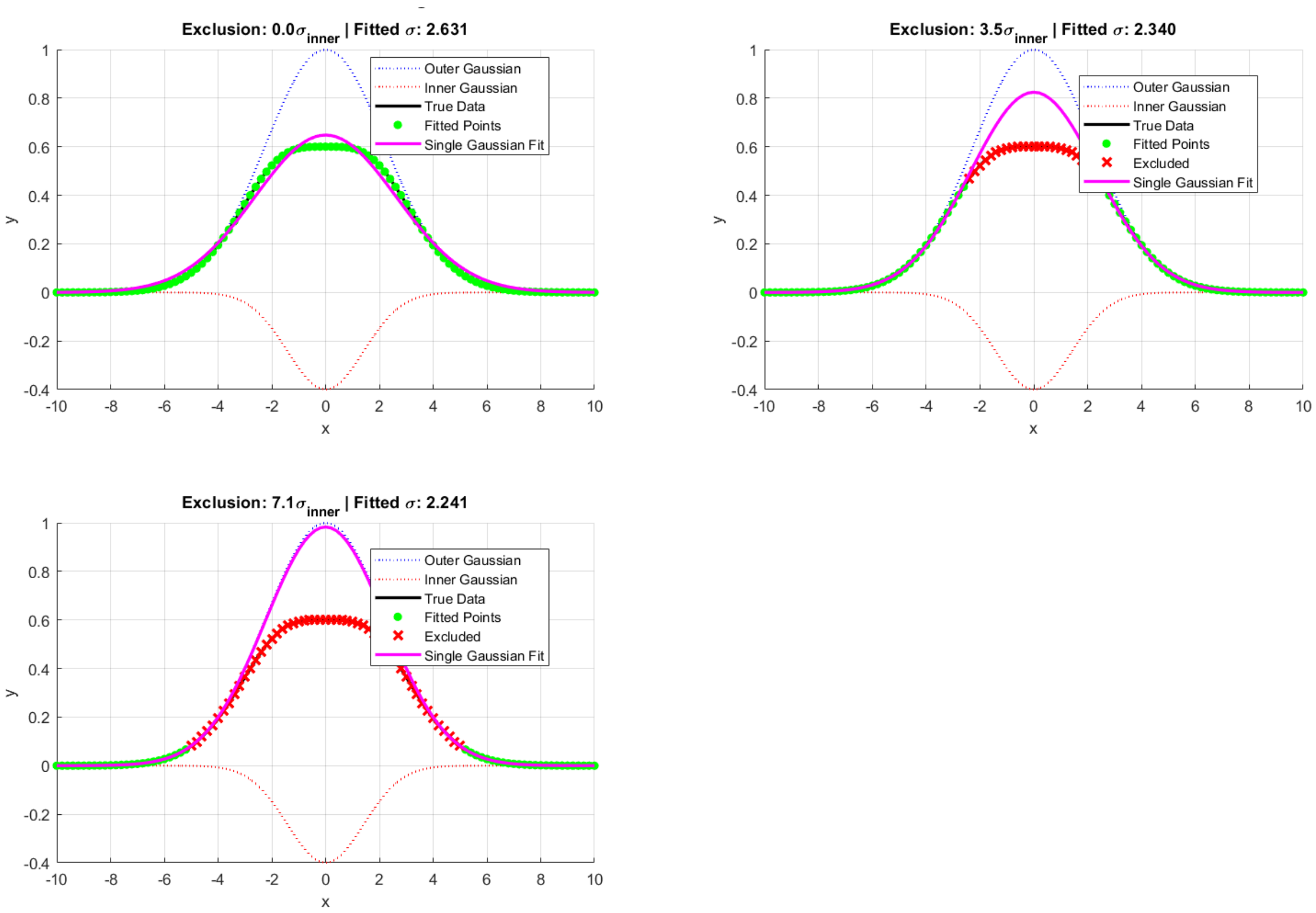

**Figure S5:** The black curve (sum of two Gaussians, data) shown in **Figure S4** is fit using Matlab's lsqnonlin to a single Gaussian in the first, second and third panels with progressively more central points excluded from the fit. The true width of the outer Gaussian (2.24) is recovered as more points are excluded, as shown systematically in **Figure S6**.

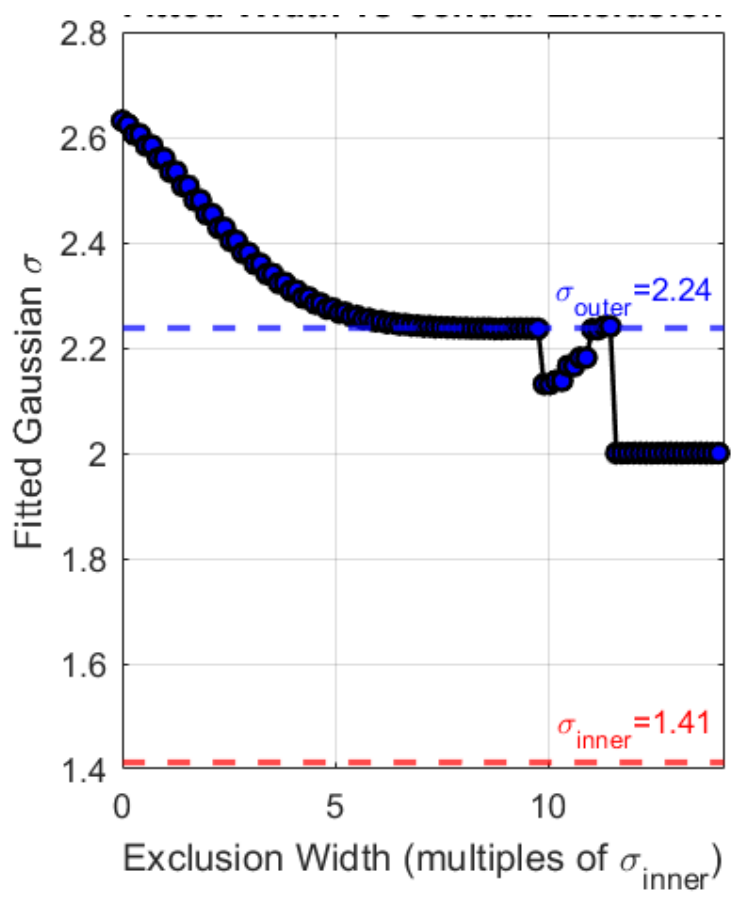

**Figure S6:** The obtained fit width from a single Gaussian fit for the sum-of-two-Gaussians curve (data) in **Figure S4** as a function of the number of points excluded from the center for the fit.

Findings from the above simple model are used for performing a two-dimensional Gaussian fit to each stroboSCAT time frame.

*Fitting: electronic evolution of stroboSCAT data:* The procedure above of observing the fit Gaussian width as a function of mask size (**Figure S6**) was used on each time delay frame of the bulk black phosphorus stroboSCAT data while performing fits to two-dimensional Gaussians:

$$S(x, y;\, t) \;=\; B(t)e^{-\frac{(x-x_0)^2}{2\sigma(t)^2_{AC}}}e^{-\frac{(y-y_0)^2}{2\sigma(t)^2_{ZZ}}}$$

*Equation S2*

where $S(x, y; t)$ is the two-dimensional spatial signal for a given time delay. $B$ is the fitted amplitude of the Gaussian and is a free parameter. Note that the amplitude $B$ is global across the two directions for a given time frame. $\sigma(t)_{AC}$ is the *1/e* width of the two-dimensional Gaussian in the AC direction and is a free parameter, and $\sigma(t)_{ZZ}$ is the *1/e* width of the two-dimensional Gaussian in the ZZ direction and is a free parameter. Fits were performed for delay times from 1.5 ps to 100 ps; longer delays are omitted due to signal-to-noise ratio limitations. For each frame and direction (AC (**Figure S7**) and ZZ (**Figure S8**)), the exclusion half-width for each species was increased until the fit width converged in spite of increasing exclusion (large-exclusion limit). Fits that became ill-defined due to excessive exclusion were not used.

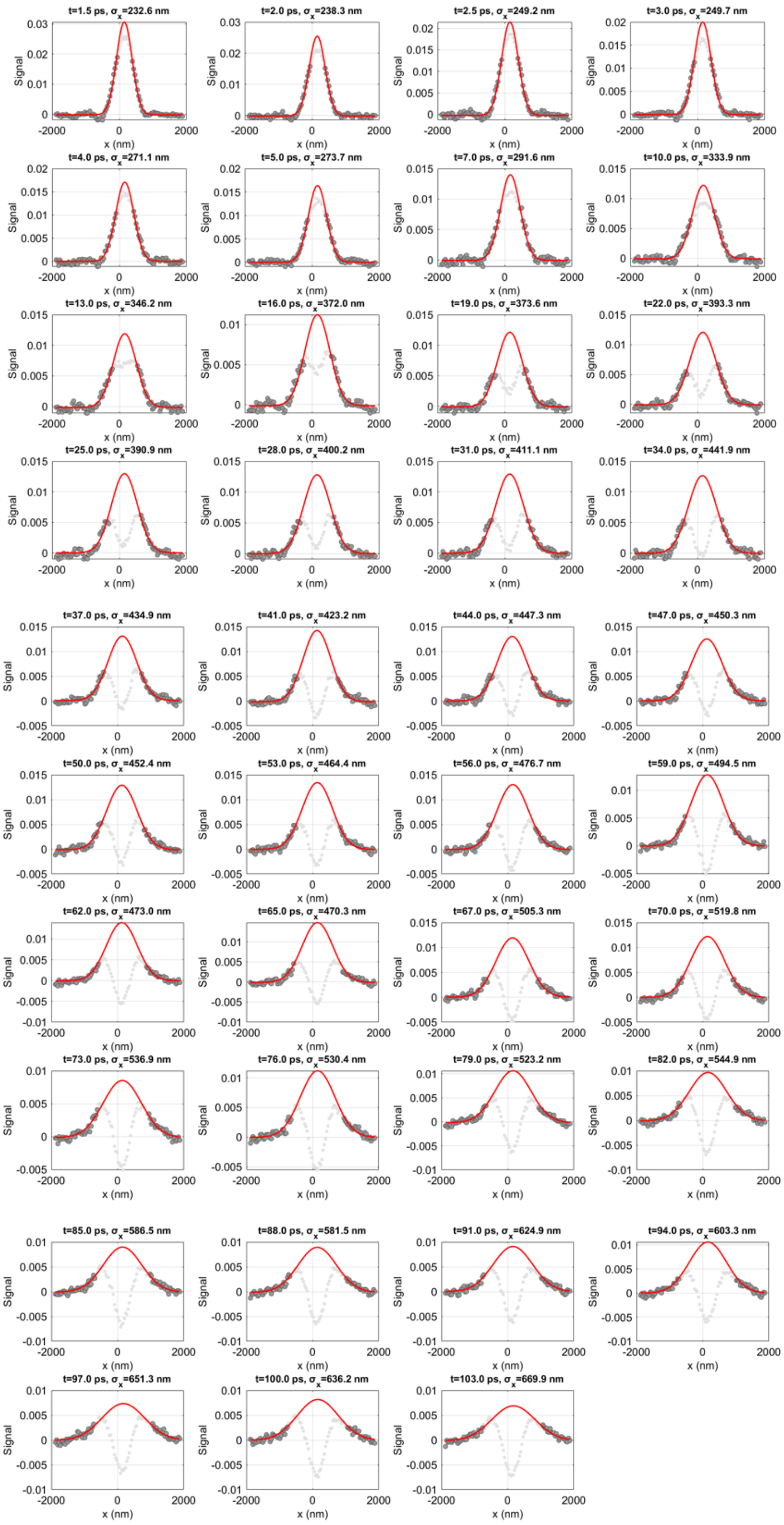

**Figure S7:** Central stroboSCAT signal cross-sections along the AC axis for each time delay for bulk black phosphorus. All data points along the cross-section are shown, but data points included for the fit are shown as dark grey open circles, and excluded data points are shown as smaller light grey points. The obtained fits are plotted in red. Delay time and fitted width are indicated above each frame.

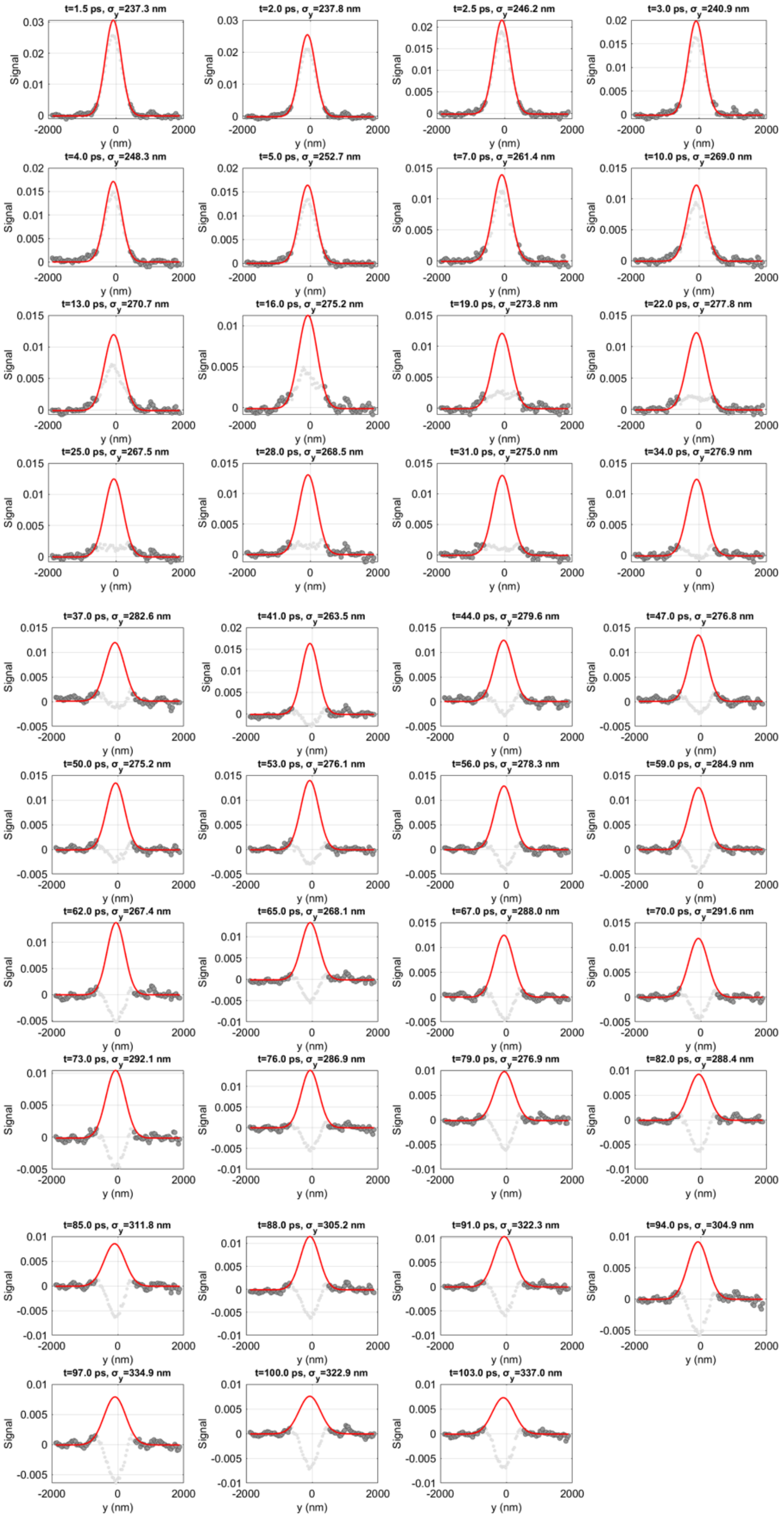

**Figure S8:** stroboSCAT signal cross-sections through the signal center along the ZZ axis for each time point for bulk black phosphorus. All data points along the cross-section are shown, but data points included for the fit are shown as dark grey open circles, and excluded data points are shown as smaller light grey points. The obtained fits are plotted in red. Delay time and fitted width are indicated above each frame.

*Mean-squared expansion and diffusivity of the electronic species in bulk black phosphorus*

The mean-squared expansion (MSE) of the electronic species in the AC direction: $\sigma_{AC}^2(t) - \sigma_{AC}^2(t_0)$ and in the ZZ direction: $\sigma_{ZZ}^2(t) - \sigma_{ZZ}^2(t_0)$ are plotted as a function of time in **Figure S9**. $t_0$ was fixed to 1.5 ps. Oscillations are observed in the MSE curves as a function of time along both directions indicating modulation of the stroboSCAT signal by acoustic modes introduced by the pump pulse, consistent with those shown in **Figure S3**. To account for these oscillations, the MSE vs time in **Figure S9** was fit to the equations:

$$\sigma_{AC}^2(t) - \sigma_{AC}^2(t_0) = 2D_{AC}t + A_{AC}cos(\omega_{AC}t + \varphi_{AC})$$

*Equation S3*

and

$$\sigma_{ZZ}^2(t) - \sigma_{ZZ}^2(t_0) = 2D_{ZZ}t + A_{ZZ}cos(\omega_{ZZ}t + \varphi_{ZZ})$$

*Equation S4*

The electronic diffusivities $D_{AC}$ and $D_{ZZ}$ were obtained from the linear (diffusive) terms in Equations S2 and S3, respectively. Here, $A_{AC}$ and $A_{ZZ}$ are the amplitudes of the oscillations, $\omega_{AC}$ and $\omega_{ZZ}$ are the frequencies of the oscillations and match well with the phonon modulations seen in the integrated stroboSCAT signal (**Figure S3**), and $\varphi_{AC}$ and $\varphi_{ZZ}$ are the phases of the oscillations. For clarity, the data shown in **Figure 3** of the main text corresponds to the data in **Figure S9** minus the sinusoidal component of the respective fits, and the linear fits in **Figure 3** correspond to the linear components of the respective fits in **Figure S9**.

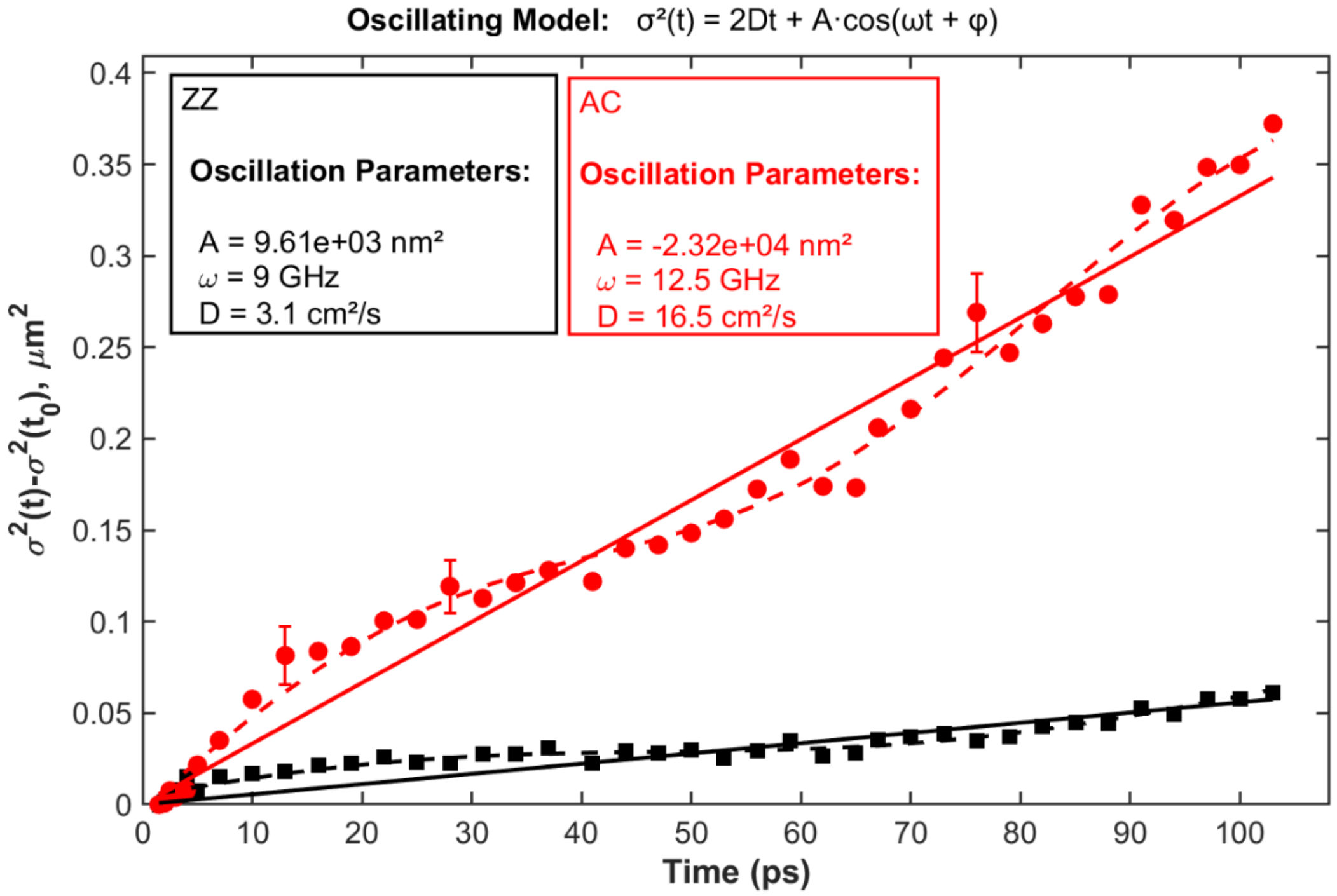

**Figure S9:** Mean-squared expansion as a function of delay time obtained from performing the masking procedure detailed above. A straight line + oscillation model fits both AC and ZZ directions. Fits suggest phonon modes of 9 GHz and 12.5 GHz modulate the signal in the ZZ and AC directions, respectively, in agreement with the Fourier transform of the data shown in **Figure S3**.

*Fitting heat diffusion of stroboSCAT data:* To obtain heat diffusivities in bulk black phosphorus, the residuals from the above electronic Gaussian fitting are fit to single negative-amplitude two-dimensional Gaussians in each frame. The mean-squared expansion of the widths in the *AC* and *ZZ* directions is calculated and diffusivities are obtained by fitting the mean squared expansion vs time to the following equations:

$$\sigma^2_{AC,heat}(t) - \sigma^2_{AC,heat}(t_0) = 2D_{AC}t$$

*Equation S5*

and

$$\sigma^2_{ZZ,heat}(t) - \sigma^2_{ZZ,heat}(t_0) = 2D_{ZZ}t.$$

*Equation S6*

Associated fits are shown in **Figures S10 and S11** and the obtained MSE curves vs. time are shown in **Figure S12**.

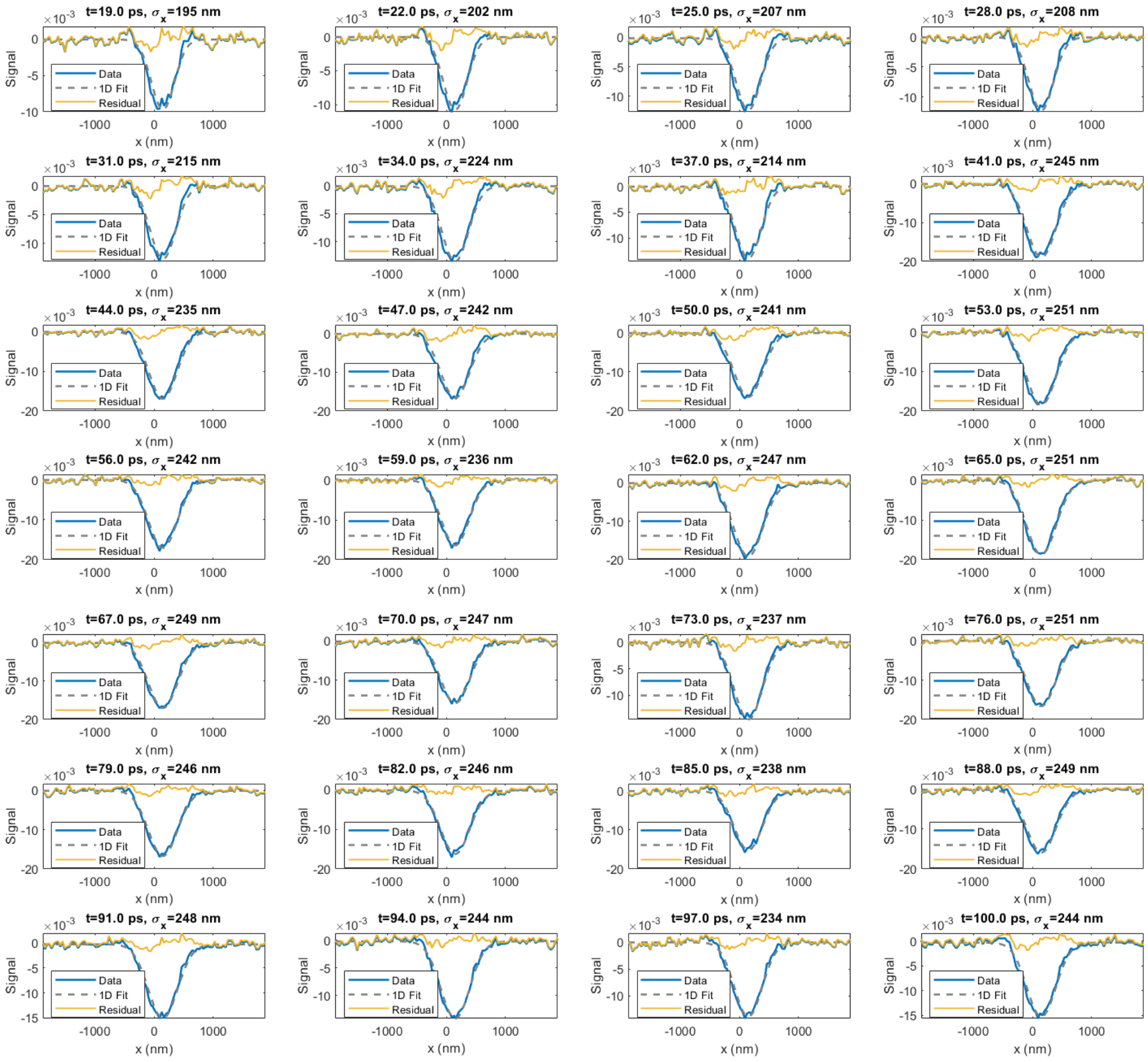

**Figure S10:** Central AC cross-sections of the remaining heat signal, following subtraction of the 2D electronic Gaussian in bulk black phosphorus (cyan); central cross-sections of the residuals of the two-dimensional Gaussian fit are shown in yellow.

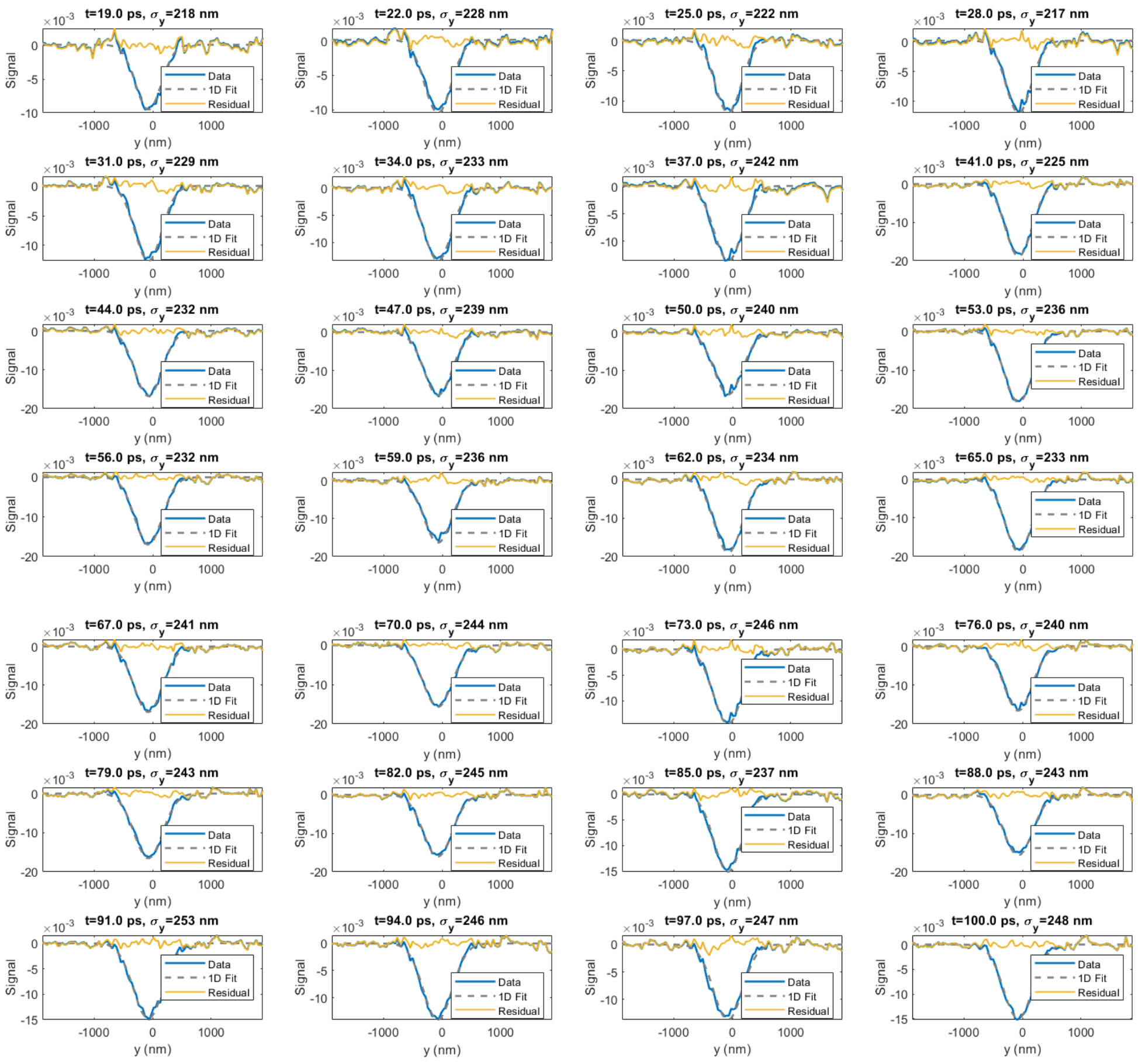

**Figure S11:** Central ZZ cross-sections of the remaining heat signal, following subtraction of the 2D electronic Gaussian in bulk black phosphorus (cyan); central cross-sections of the residuals of the two-dimensional Gaussian fit are shown in yellow.

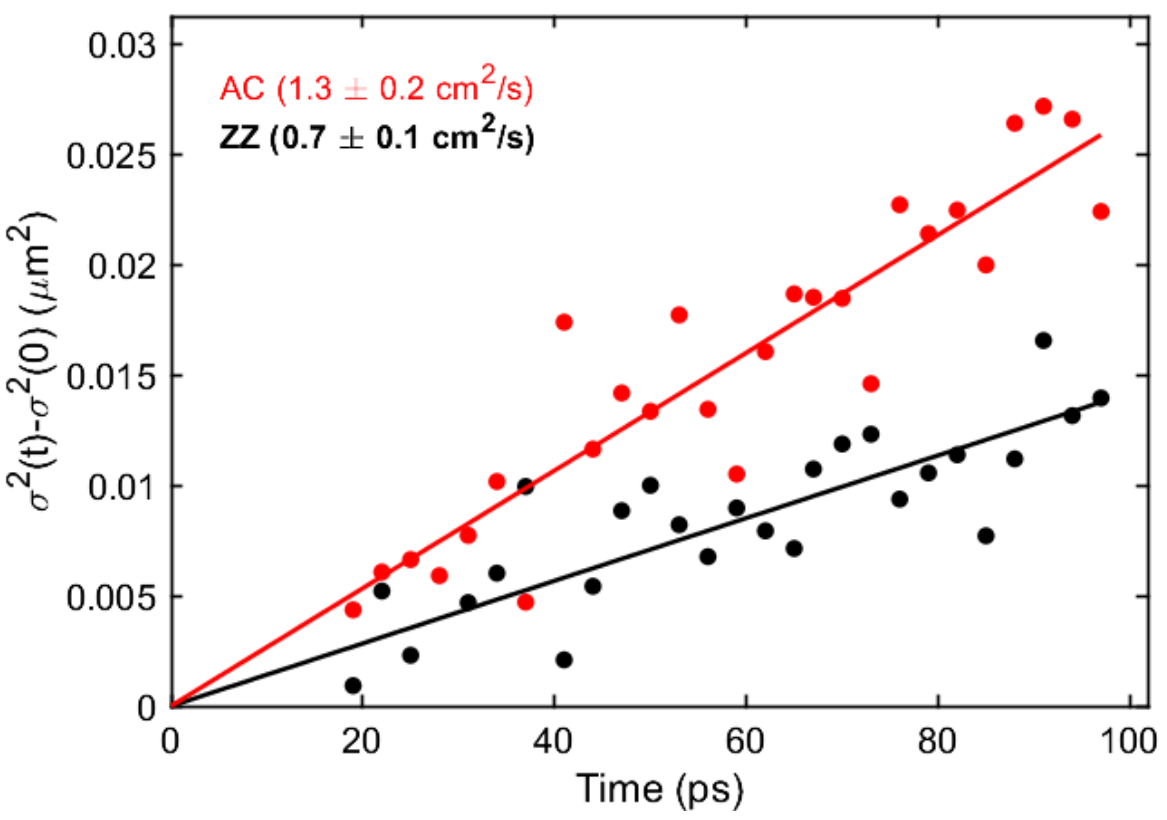

**Figure S12:** Mean-squared expansion of the two-dimensional Gaussian fits (**Figures S10 and S11**) to the residuals of the *electronic* Gaussian fitting described above provide *heat* diffusivity for bulk black phosphorus along the AC and ZZ directions.

**Monolayer identification and data analysis**

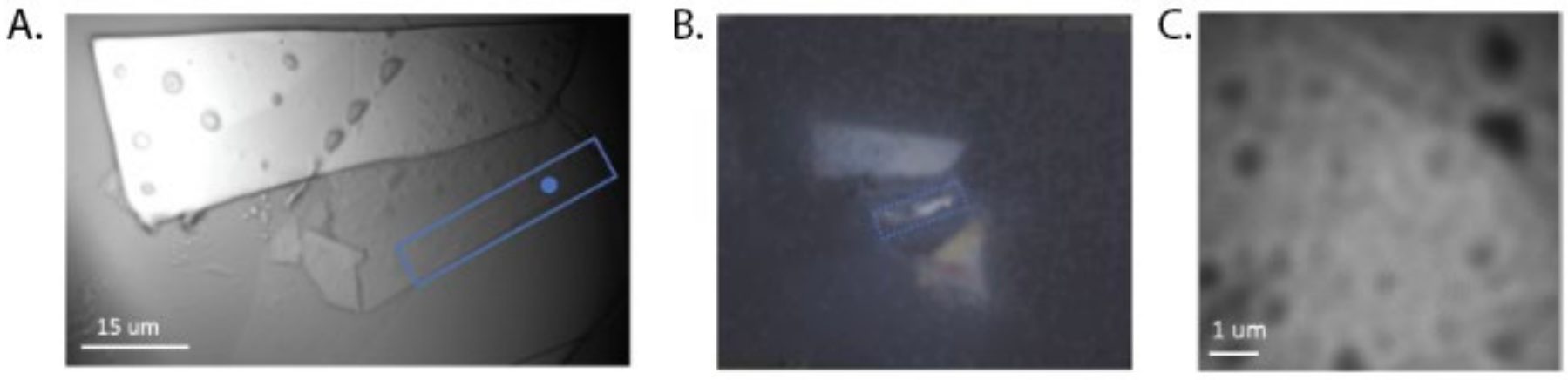

**Figure S13**: Monolayer identification. (A) Reflection image of black phosphorus flakes. (B) White light image overlaid with monolayer emission image. Blue box highlights monolayer region as defined by emission image and corresponds to blue box in panel (A). (C) Ground state reflectance image of monolayer region in which the stroboSCAT measurements in **Figure 4** were performed. Location shown with blue dot in panel (A).

*Monolayer diffusivity analysis.* For the monolayer data, the large-exclusion limit is not attained in the way shown for the bulk data in **Figure S5** due to lower signal-to-noise ratio; therefore, a frame-by-frame fitting protocol was used to obtain monolayer diffusivities. For this dataset, fitting identical to the bulk dataset is performed. A two-dimensional spatial Gaussian fit is used for each time delay. The Gaussian center is determined from the centroid of early frame fits without masking (when heat is minimal) and is held fixed thereafter, while the mask center is determined by fitting the late-time residual structure when heat is most prominent to a single two-dimensional Gaussian. We observe a significant offset (~50 nm) between the heat and electronic signal centers. For each frame, a sequence of mask exclusion radii was tested; only the Gaussian amplitude and widths were optimized, while pixels within the exclusion radius were removed. The mask width is taken from the longest stability plateau in the mask sweep (or, if no stable plateau was identified, then the largest exclusion providing non-NaN fits). The squared $1/e$ half-widths are then converted into mean-squared expansion by referencing to the 1.5 ps time delay width, and the diffusivity is extracted from linear fits of the function

$$\sigma_{AC}^2(t) - \sigma_{AC}^2(t_0) = 2D_{AC}t$$

*Equation S7*

and

$$\sigma_{ZZ}^2(t) - \sigma_{ZZ}^2(t_0) = 2D_{ZZ}t.$$

*Equation S8*

This protocol is used to minimize uncertainty in diffusivity values due to exclusion choice systematics. Cross-sections of fits to each time frame are shown in **Figures S14 and S15**. Despite lower signal-to-noise ratio in particular in the ZZ direction in monolayer BP, the peak of its cross-section is constrained by the fact that we fit to a 2D Gaussian. As a result, the data along the stronger AC direction aids in constraining the fit width along ZZ.

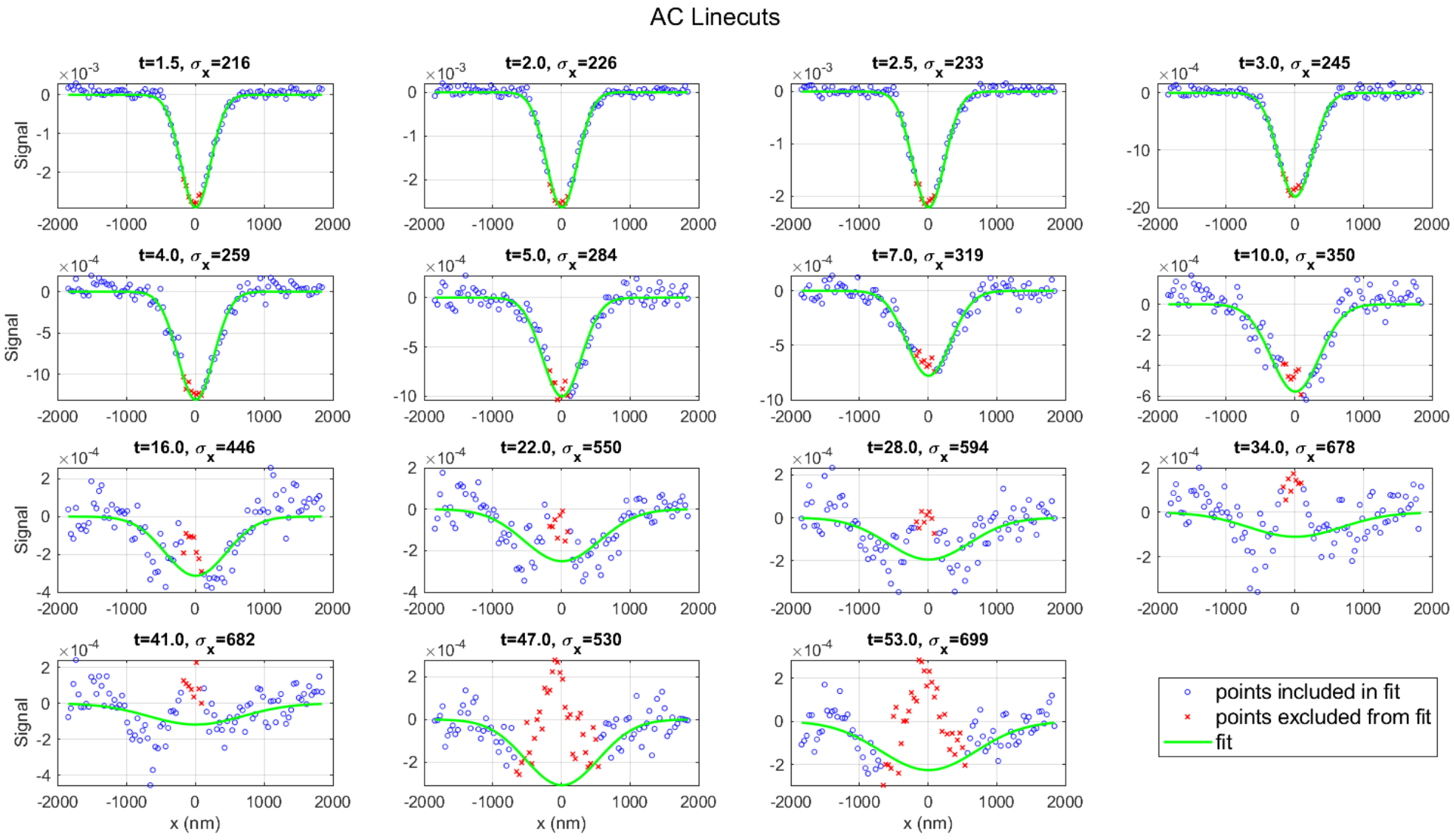

**Figure S14:** Central stroboSCAT signal cross-sections along the AC axis for each analyzed time delay for monolayer black phosphorus. All data points along the cross-section are shown. Data points included for the fit are shown as blue circles; excluded data points are shown as red crosses; fit is shown in green. Delay time and fitted width are shown above each frame.

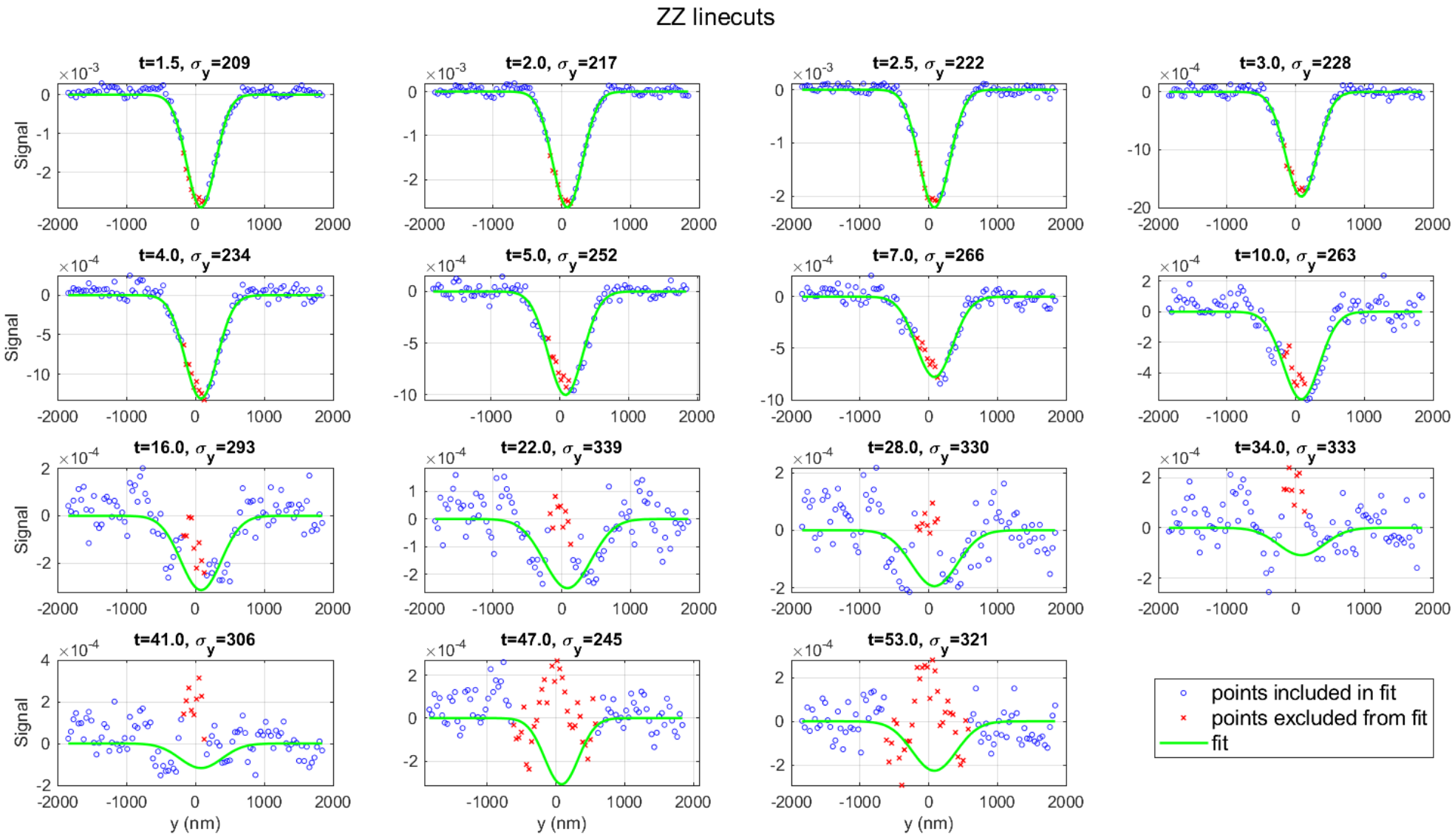

**Figure S15:** Central stroboSCAT signal cross-sections along the ZZ axis for each analyzed time delay for monolayer black phosphorus. All data points along the cross-section are shown. Data points included for the fit are shown as blue circles; excluded data points are shown as red crosses; fit is shown in green. Delay time and fitted width are shown above each frame.

**Supporting References**